\documentclass[10pt,letterpaper]{article}
\usepackage{opex3}
\usepackage{cite}
\begin{document}

%%%%%%%%%%%%%%%%%% title page information %%%%%%%%%%%%%%%%%%
\title{Spatially resolved pump-probe study of single-layer graphene produced by chemical vapor deposition}

\author{Brian A. Ruzicka,$^1$ Shui Wang,$^2$ Jianwei Liu,$^1$ Kian-Ping Loh,$^2$ Judy Z. Wu,$^1$ and Hui Zhao$^{1,*}$}

\address{$^1$Department of Physics and Astronomy, The University of Kansas, Lawrence, Kansas 66045, USA\\
 $^2$Department of Chemistry, National University of Singapore, 3 Science Drive 3, Singapore 1175436}

\email{*huizhao@ku.edu} %% email address is required


\begin{abstract}
Carrier dynamics in single-layer graphene grown by chemical vapor deposition (CVD) is studied using spatially and temporally resolved pump-probe spectroscopy by measuring both differential transmission and differential reflection.  By studying the expansion of a Gaussian spatial profile of carriers excited by a 1500-nm pump pulse with a 1761-nm probe pulse, we observe a diffusion of hot carriers of 5500~cm$^{2}$/s.  We also observe that the expansion of the carrier density profile decreases to a slow rate within 1 ps, which is unexpected.  Furthermore, by using an 810-nm probe pulse we observe that both the differential transmission and reflection change signs, but also that this sign change can be permanently removed by exposure of the graphene to femtosecond laser pulses of relatively high fluence.  This indicates that the differential transmission and reflection at later times may not be directly caused by carriers, but may be from some residue material from the sample fabrication or transfer process.
\end{abstract}

 % REPLACE WITH CORRECT OCIS CODES FOR YOUR ARTICLE

Graphene has been promoted as an excellent material for use in the fabrication of high speed transistors due to its large electron mobility, despite of the
lack of a band gap \cite{s306666,n47274}.  In addition to the applications of graphene, it is also a very interesting material to study in terms of fundamental physics, as it is one of very few easily accessible, truly two-dimensional systems.  As such, several different types of graphene have been studied extensively using ultrafast pump-probe techniques, including mechanically exfoliated graphene \cite{ox172326,apl97221904,b83153410,b83121404}, epitaxial graphene on silicon carbide (SiC) \cite{apl92042116,nl84248,l101157402,apl94172102,b80245415,nl101308,l104136802,apl96081917,b82195414}, graphene oxide \cite{apl98141103,apl98121905,jpcc11519110,jpcl21972}, reduced graphene oxide \cite{apl95191911,apl96173106,jap109084322,apl98121905,jpcc11519110,b82195414},
and graphene produced by chemical vapor deposition (CVD) \cite{apl97163103,apl96081917,nl113184,c494781,acsnano53278,nl111540}. In these studies, a pump pulse is used to inject charge carriers by exciting electrons from the valence band to the conduction band. Since these carriers occupy some states in the conduction band, absorption of a time-delayed probe pulse is reduced due to phase state filling effect. Such a saturable absorption can be used in nonlinear optical applications \cite{afm193077}. Owing to this effect, the differential transmission and/or reflection of the probe pulse monitors the dynamics of these carriers. These studies can provide information about the transport of carriers \cite{b82195414}, energy relaxation of carriers \cite{apl92042116}, phonon dynamics and the related substrate effects \cite{apl96081917,b83121404,nl113184}, and Fermi energy of layers in graphene samples \cite{l104136802}, just to name a few. However, several questions about the mechanism of this technique still remain open. For example, it is still unclear as to the exact source of the signal at later time delays, and the influence of the ultrafast laser pulses on the samples is still in debate. Furthermore, most studies are limited to measurements where the pump and the probe spots overlap in space, while spatially resolved studies are relatively few in number.

Graphene produced by CVD has been the focus of recent studies since it is easily scalable for growth on Si substrates \cite{s3241312}, and since high frequency transistors based on this type of graphene have been demonstrated \cite{n47274}. Here we present a unique, spatially and temporally resolved pump-probe study of charge carrier dynamics in CVD graphene. We temporally and spatially resolve the carrier dynamics by measurement of the differential transmission and differential reflection.  Using these we are able to monitor the expansion of the carrier density profile over time and observe the diffusion of carriers.  We measure a carrier diffusion coefficient of 5500~cm$^{2}$/s and also observe that the expansion of the carrier density profile slows down substantially after about 1~ps.  Additionally, we observe that, depending on the probing wavelength used, both the differential transmission and differential reflection will change signs.  This sign change can be permanently removed by exposure of the graphene to femtosecond laser pulses of relatively high fluence.  This permanent modification may indicate that the differential transmission and differential reflection at later time delays is caused by some residue material from the sample fabrication or transfer process, rather than carriers.

\begin{figure}
 \centering
 \includegraphics[width=12cm]{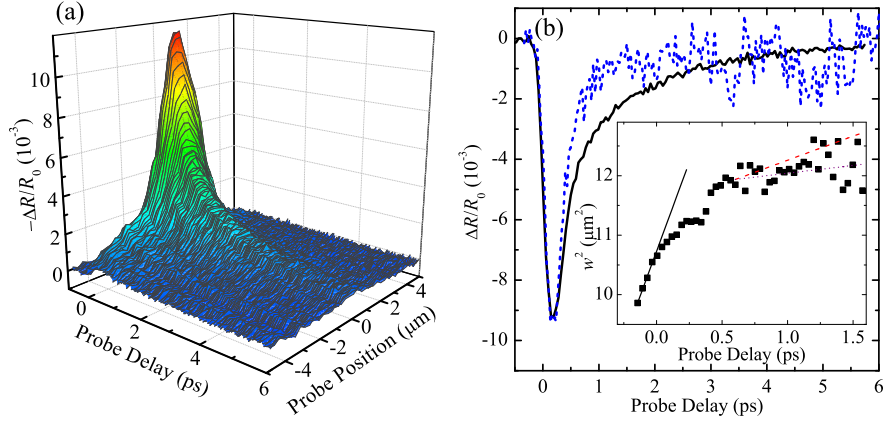}
 \caption{(a) Differential reflection of the CVD graphene sample as a function of the probe delay and probe position, measured with a 1500-nm pump and a 1761-nm probe. (b) Black solid line: the differential reflection as a function of the probe delay when the center of the probe spot overlaps with the pump spot. Blue dotted line: same measurement but performed as the sample has been exposed to a laser pulse  of 2.3 mJ/cm$^2$ for 20 minutes. The squares in the inset of (b) shows the squared width as a function of probe delay, deduced from Gaussian fits to the profiles shown in (a).}
  \label{diffusion}
\end{figure}

The single-layer graphene samples were grown by CVD, as described by Wang {\it et. al.} \cite{apl95063302} and subsequently transferred to quartz substrates. The experimental setup is similar to that described in previous experiments performed by the authors \cite{apl92112104,b82195414,apl97262119}. Carriers are injected by a tightly focused pump pulse with a central wavelength of 1500 nm, and detected by a tightly focused, time delayed, and spatially scanned probe pulse of 1761 nm. The pump and the probe pulses are obtained from the signal and the idler outputs, respectively, of an optical parametric oscillator that is pumped by a Ti:sapphire laser at 80 MHz.  The cross-correlation between pump and probe is approximately 250~fs. Differential reflection of the probe, defined as $\Delta R / R_0 = (R-R_0)/R_0$, where $R$ and $R_0$ are the reflections with and without the presence of the pump pulse, is measured as a function of probe delay and probe position, as shown in Fig.~\ref{diffusion}(a) (plotted as $- \Delta R/R_0$). The pump fluence is 0.5 mJ/cm$^2$. The signal is negative at all probe delays and probe positions. It decays rapidly with time and the spatial profile broadens. The solid curve in Fig.~\ref{diffusion}(b) shows the signal as a function of the probe delay when the center of the probe spot overlaps with the pump spot (defined as zero probe position). We can fit its decay with a bi-exponential decaying function and we find a fast time constant of 0.2~ps, which can be attributed to carrier thermalization, and slower one of 1.6~ps. 

The spatially resolved measurement allows us to obtain the diffusion coefficient of the carriers. The spatial profile of the signal remains Gaussian and broadens as time increases, owing to the carrier diffusion. Since the initial spatial distribution of the carriers is Gaussian due to the Gaussian shape of the laser spots, carrier diffusion will cause the distribution to remain Gaussian, with the square of the full width at half maximum (FWHM) of the profile increasing linearly with time: $w^{2}(t) = w_0^2 + 16 \ln (2) D t$, where $w_0$ is the initial width and $D$ is the carrier diffusion coefficient \cite{b79115321}. The symbols in the inset of Fig.~\ref{diffusion}(b) show the squared width deduced from Gaussian fits to the measured profiles as a function of the probe delay. From a linear fit to the data points at early probe delays [solid line in the inset of Fig.~\ref{diffusion}(b)], we deduce a diffusion coefficient of 5500$\pm$400~cm$^2$/s. The diffusion coefficient is related to the mobility by Einstein relation, which is, for graphene with a linear dispersion, $D= \mu k_{B}T/2q$, where $k_{B}$ is Boltzmann's constant, $T$ is the carrier temperature, and $q$ is the electron charge. The carriers are excited with an excess energy of 0.413 eV.  Since carriers will rapidly redistribute their energy and form a Fermi-Dirac distribution, this corresponds such a distribution with a temperature of $T = 2200$~K. From these values, we obtain a mobility of $6 \times 10^{4}$~cm$^2$/Vs.

The mobility we deduced by the optical measurement is higher than most reported values from electrical measurements (e.g. 2000, 3700, and 16000~cm$^2$/Vs).\cite{jap107044310,n457706,nl104328}. One big challenge of CVD graphene growth and application is the formation of grain boundaries. These discontinuities of the graphene film cause the conductivity to decrease. Electrical measurements performed on large graphene samples are influenced by these grain boundaries. Our optical measurement is performed locally over a length scale of about only 5~$\mu$m, close to the grain size of the samples.  Therefore, we view the results as intrinsic mobility of graphene that is not influenced by grain boundaries. Indeed, the value we deduce is comparable to that of other types of graphene samples.  Additionally, we note that the carrier temperature is changing during the measurement due to cooling, and therefore is not exactly known.  However, the calculation serves well for an order of magnitude estimate of the mobility.

The sub-linear expansion shown in Fig.~\ref{diffusion}(b) indicates that the diffusion coefficient decreases with time. At later time delays, when the expansion of the spatial profiles appears to slow ($> 1$~ps), we would obtain a diffusion coefficient of 250$\pm$140~cm$^2$/s by a fit [blue dotted line in the inset of Fig.~\ref{diffusion}(b)]. Previous electrical measurements have shown that the carrier mobility in graphene is only weakly dependent on the carrier temperature \cite{s3121191}. Hence, from the Einstein relation, carrier cooling is expected to cause the diffusion to slow down. We can check if the value at later time delays is reasonable, based on the decrease of the temperature from its initial value and the diffusion coefficient deduced from early time delays. Assuming the carriers have reached thermal equilibrium with the lattice at this time and are at a temperature of 293~K, the diffusion coefficient would have decreased by a factor of about 7.5 from its initial temperature of about 2200~K (0.413~eV).  Since the initial diffusion coefficient is 5500~cm$^{2}$/s, the diffusion coefficient at late time delays should be about 730~cm$^{2}$/s.  The solid red line in Fig.~\ref{diffusion} indicates the expansion that would be expected if the carrier temperature is decreased to 295~K. This most likely overestimates the decrease in the carrier temperature and yet is still larger than what we have observed.

Since the carrier density profile should expand at a much faster rate than observed, we suspect that the differential transmission is caused by some additional mechanism besides the charge carriers. In this scenario, the pump pulse would excite some foreign agents that exist on graphene, or cause a local hot spot. The spatial distribution of such excitations is determined by the shape of the pump spot. It would not expand at all or would expand at a rate much smaller than the mobile carriers. The measured profile is composed of two profiles, one caused by the carriers and the other by this additional contribution. At early time delays, the carrier contribution dominates. Hence the width measured accurately reflects the carrier dynamics, and we can still measure a diffusion coefficient. Since the carrier contribution decreases rapidly with time, at late delays the thinner spatial profile of the additional mechanism start to dominate the measured width, causing an apparent slow expansion of the measured profile.

The ultraslow or non-expansion of the profile of the differential transmission signal observed in this spatially resolved measurement can also provide insight to an open issue in ultrafast optical studies of carrier dynamics in graphene. So far, the mechanism that is responsible for the differential reflection/transmission at later time delays, i.e. time delays beyond roughly 1~ps, is under debate, due to inconsistencies in the sign.  Specifically, at later time delays
(beyond 0.3 to 1~ps) it has been observed that the differential transmission will undergo a sign flip from positive to
negative, in epitaxial graphene on SiC \cite{l104136802,l101157402}, stacked graphene films on quartz produced by low-pressure CVD \cite{apl97163103}, exfoliated graphene on mica substrates \cite{b83153410}, and biased graphene oxide devices on glass \cite{apl98141103}. This indicates that the dominating contribution has changed.  In other studies of graphene of various types on various substrates however, no sign flip was observed \cite{b83153410,apl94172102,apl92042116,nl84248,b83121404,nl101308,apl95191911,apl97221904,ox172326,b80245415,apl96173106,b82195414,l104136802,l101157402,apl96081917,jap109084322,apl98121905,apl97141910,nl113184}.  This sign flip has been assigned to various physical mechanisms including a negative differential transmission induced by the doping of layers in epitaxial graphene on SiC\cite{l104136802,l101157402} and lattice heating effects \cite{apl97163103}.  Additionally, if the differential reflection or transmission is caused by something other than carriers, then pump-probe studies cannot strictly be used to observe the energy relaxation of carriers, since the probe may not be sensitive to carriers alone. Hence, investigation into the source of such a sign flip is necessary.

A sign flip can be induced by the superposition of two mechanisms that cause differential transmission/reflection signals of opposite signs. This is consistent with our observation, where two mechanisms contributing to the signal have very different transport behaviors. However, they have the same sign, and hence no sign flip. Unlike graphene with zero bandgap, most materials have a finite bandgap, and hence the sign of a differential transmission signal can be wavelength dependent. Based on this, we repeat the measurement by changing the pump/probe wavelength from 1500/1761 nm to 750/810 nm. The 750-nm pump pulse is obtained by second-harmonic generation of the signal output of the optical parametric oscillator of 1500 nm. The 810-nm probe is obtained directly from the output of the Ti:sapphire laser that pumps the optical parametric oscillator. In order to obtain more information about the sign flip, in this experiment, we measure both the differential reflection and the differential transmission.

\begin{figure}
 \centering
 \includegraphics[width=12cm]{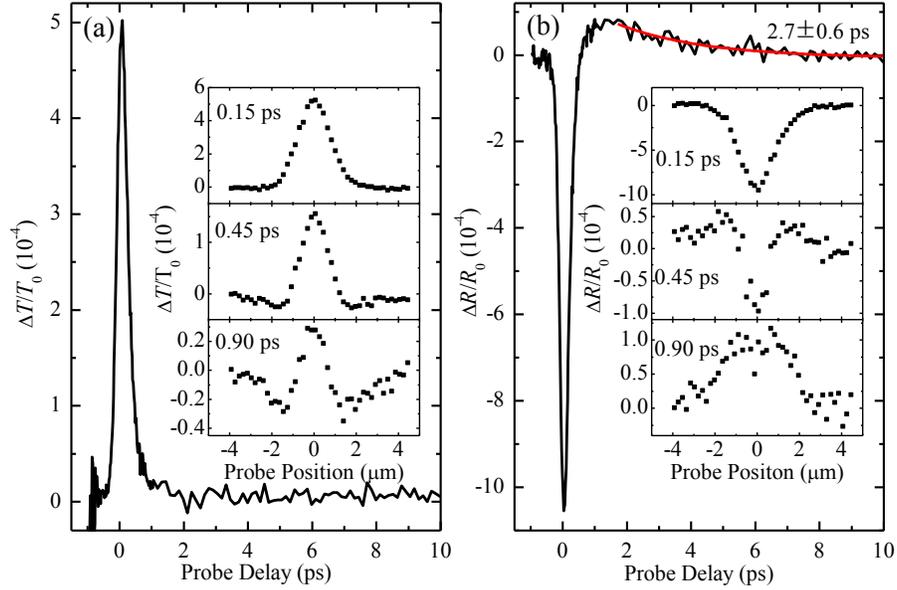}
 \caption{The differential transmission (a) and reflection (b) signals measured for
 a 750-nm pump fluence of 0.7~mJ/cm$^{2}$ with an 810~nm probe. The insets show the spatial profiles of the signal at several probe delays measured by scanning the probe spot across the pump spot.}
   \label{signflip}
\end{figure}

Figures~\ref{signflip}(a) and \ref{signflip}(b) show the differential transmission and reflection, respectively, measured for a pump fluence of 0.7~mJ/cm$^{2}$ and with the pump and probe spots overlapped. At early delays, the differential transmission is positive and the differential reflection is negative. This is consistent with Fig.~\ref{diffusion} and previous observations, showing that charge carriers in graphene cause a reduction of the absorption. The differential transmission decays much faster than that measured with the 1500-nm probe. This is also consistent with the fact that the 1500-nm pulse probes lower energy states. However, at later delays, while the differential transmission remains positive, the differential reflection changes sign after approximately 530~fs. The positive signal decays with a much longer time constant of 2.7$\pm$0.6~ps, and during this period of time, no differential transmission signal is seen. Since the carrier-dominated signals of differential transmission and reflection seen at early delays are similar, these two observations strongly suggest that the positive differential reflection signal is caused by something other than the carriers.  Although no sign flip can be clearly seen for the differential transmission measurement, as we look to the sides of the spot, by scanning the probe along the x-direction, as shown in the insets, we see that the sign of the differential transmission does in fact become negative at later time delays: at 0.45 ps, negative wings can be seen, which become comparable to the main peak at 0.9 ps. Meanwhile, the spatial profiles of the differential reflection signal [inset of Fig.~\ref{signflip}(b)] show that the profile changes from a positive Gaussian to a negative Gaussian (at 0.9 ps), with a superposition of the two at 0.45 ps. Hence, this additional contribution does exist in both differential transmission and differential reflection. The absence of the sign flip in the differential transmission measured at $x=0$ is simply due to the fact that this additional contribution does not win over the carrier-induced differential transmission at $x=0$.

\begin{figure}
 \centering
 \includegraphics[width=8.5cm]{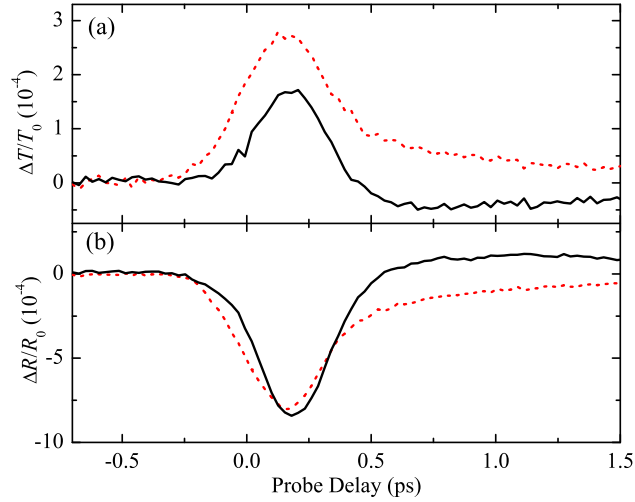}
 \caption{The differential transmission (a) and reflection (b) signals for a pump fluence of
 0.5~mJ/cm$^{2}$ measured before (black solid line) and after (red dashed line)
 20-minute exposure to a 1500-nm and 2.3~mJ/cm$^{2}$ laser pulse.}
 \label{modification}
\end{figure}

In an effort to further confirm our interpretation and to eliminate such a contribution, we move the probe spot to the side at about $x=1.5$~$\mu$m, and measure the signals as functions of probe delay. The results are plotted as the block solid lines in Fig.~\ref{modification} Both curves display a sign flip. We then expose the sample to a 100-fs, 1500-nm pulse with a 2.3-mJ/cm$^2$ fluence and an 80-MHz repetition rate for 20 minutes. After the exposure, we measure the differential transmission and reflection again, as shown as the red dotted lines and we no longer see the sign flip in either curves. Apparently, the foreign agent has either been removed or is no longer sensed by the pump and/or probe pulses after the exposure.  We also note that this behavior was observed at several locations throughout the sample.

Such an observation encourages us to revisit the 1500/1760-nm pump/probe scheme, where we have seen a slow decay component with little or no spatial expansion (Fig.~\ref{diffusion}). We repeat the time scan with $x=0$ after the sample was exposed to 1500-nm pulse with the same conditions, as shown as the blue dotted line in Fig.~\ref{diffusion}(b). Clearly, the slow component was removed. Furthermore, this additional contribution is of the same sign as carrier-induced signals with the 1761-nm probe. Hence, this agent causes photo-bleaching for the 1761-nm probe, but photo-induced absorption for the 810-nm probe. Unfortunately, we could not repeat the diffusion measurement since we could not uniformly illuminate a large area with sufficient fluence. Nevertheless, it is clear that the late delay signal is dominated by this foreign agent, which prevented us from measuring diffusion at late delays.

It is unlikely that the graphene is being destroyed or modified during exposure to the high intensity laser pulses, since the temporal dynamics of the differential transmission and reflection signals remain graphene-like (i.e. the signal decays on a time scale on the order of hundreds of femtoseconds). Recently, it was observed that the exposure of graphene to femtosecond laser pulses of high fluences that are still below 200~mJ/cm$^{2}$ does not destroy the graphene, but instead leads to the formation of defects, i.e. an increased number of grain boundaries \cite{apl99051912}.  On the other hand, Currie {\it et al} found that structural modification of CVD graphene by femtosecond laser pulses did not occur for optical fluences below 14~mJ/cm$^2$ \cite{apl99211909}.  While the fluences used here are still below this threshold, it is possible that the structure of the graphene is being modified by the laser, if these particular samples have a lower damage threshold.  This would then imply that signal at later time delays is caused by an intrinsic property of graphene.  However, this is unlikely for three reasons: the signal that we measured changes sign for the 810~nm probe; a signal with no sign flip has been observed in other CVD graphene samples, reduced graphene oxide and epitaxial graphene samples on SiC \cite{apl96173106,b82195414,jap109084322}; and the sign of the signal at later time delays appears to be wavelength dependent.

Considering the three observations discussed here (expansion of the differential transmission/reflection profile due to carrier diffusion, the presence of the
sign change, and the modification of the differential signal caused by the high intensity laser pulses), we can draw a few possible conclusions about the differential
transmission/reflection signals.  First, the signal at later time delays may not be directly due to carriers.  Second, this signal - the sign of which is wavelength dependent - is not an intrinsic property of graphene.  Finally, this component of the signal can be removed from our samples after exposure to a relatively high laser fluence.  One possibility, which will require further study, is that the secondary component of the signal is due to something from either the CVD growth process or from the transfer process of the graphene from the growth substrate to glass, as described in Ref.~29.  This would explain why it is difficult to predict why it shows up in different types of samples, and in this case, can be removed by exposure to moderate laser fluence. Recently, the source of the sign flip have been attributed to various mechanisms. Although these explanations are reasonable, our results suggest that such assignments can only be made after exclusion of foreign effects such as what was discovered here.  

In summary, we have simultaneously measured the temporally and spatially resolved differential reflection and differential transmission of CVD graphene.
We successfully measured a diffusion coefficient of hot carriers of 5500~cm$^2$/s by using a 1761-nm probe pulse. The expansion of the signal profile decreases to an unexpectedly slow rate after about 1~ps.  We also observe a flip in the sign of both differential transmission and reflection when an 810-nm pulse is used as the probe, which can be permanently removed by exposure of the graphene to femtosecond laser pulses of a relatively high fluence.  Based on these observations, we conclude that the differential reflection and differential transmission at later time delays may not be directly due to the carriers, but could be caused by residue material from the sample fabrication or transfer process.  This may explain the observation of such a signal in recent ultrafast optical studies of graphene.

\section*{Acknowledgments}

We acknowledge support from the US National Science Foundation under Awards No. DMR-0954486 and No. EPS-0903806, and matching support from the State of Kansas through Kansas Technology Enterprise Corporation.

\end{document}